# Probing the temperature dependence of dielectric function of ternary transition metal dichalcogenides: towards thermo-driven ultrathin photonic components

*A. V. Margaryan[1, +], M. A. Levonyan[1, +], M. L. Sargsyan[1, +], D. A. Karakhanyan[1], M. H. Hayrapetyan[1], K. S. Novoselov[2, 3] & D. A. Ghazaryan[1, *]*

*[1]Laboratory of Advanced Functional Materials, Yerevan State University, Yerevan, 0025, Republic of Armenia*

*[2]Institute for Functional Intelligent Materials, National University of Singapore, Singapore, 117575, Republic of Singapore*

*[3]Department of Materials Science and Engineering, National University of Singapore, Singapore, 117575, Republic of Singapore*

*[+]These authors contributed equally to this work*

*[*]The correspondence should be addressed to: davitghazaryan@ysu.am*

**Transition metal dichalcogenides (TMDs), along with their ternary derivatives, have attracted considerable attention mostly due to pronounced excitonic resonances emerging in visible (Vis) and near-infrared (NIR) spectral regions, enabling strong light–matter interaction. Nevertheless, a comprehensive insight of the temperature-dependent optical dispersions for the most of representatives of the family remains yet unrevealed. Here, we report on systematic studies of dielectric permittivity functions of uniaxial ternary MoSSe and WSSe across 430-1000 nm spectral region over a broad temperature window of 80–670 K. We show that the temperature evolution of their dielectric responses is governed by Varshni's formalism in Vis spectral region further affecting their high refractive index properties at the lossless NIR spectral tails. Furthermore, we exploit the measured optical dispersion of ternary WSSe designing ultrathin plano-convex NIR photonic lenses that demonstrate continuous modulation of performance with temperature variation. Our work provides critical insights for the creation of next-generation thermo-driven nanophotonic and optoelectronic devices.**

## Introduction

TMDs[1–4] have attracted considerable attention as a versatile class of next-generation semiconductors recently reemerged in two-dimensional (2D) form[5–7] following the isolation of graphene[8,9]. Nowadays, they constitute an eminent family of van der Waals (vdW) crystals[10,11] with more than a three dozens of representatives with crystalline structures in a variety of polymorphic phases[12–14], all composed of covalently bonded atomic layers weakly coupled by vdW interaction. This vdW family comprises uniaxial and biaxial crystals displaying pronounced dielectric responses with refractive indices[15–20] surpassing traditional Si and Ge across Vis and NIR spectral regions[21,22] in the case of former, also possessing in-plane birefringences higher than traditional rutile[23], in the case of latter[24,25]. Some of the defining microscopic features of these TMD crystals are thickness dictated indirect-*to*-direct bandgap transitions[26] and strong excitonic effects accompanied by anomalously high binding energies[27–29] arising from the reduced dielectric screening and quantum confinement allowing a Bohr radii of an order of 1 nm, leading to enhanced light–matter interaction. Furthermore, a variety of their derivatives, such as ternary alloys[30,31] or Janus structures[32,33], along with a recent addition of stacking-order sensitive sliding ferroelectrics[33–36] expand their versatility in terms of tuneability or polarization switching[37–39] of the excitonic states and emergence of non-linear optical effects[36,40], especially pronounced in atomically thin layers. These and many other intriguing properties position TMD crystals, as well as their ternary alloys, into the list of particularly prospective candidates for further advancement[41,42] of nanophotonic and nanoscale optoelectronic devices, such as ultrathin quarter waveplates[43], ultraconfined exciton-polaritonic waveguides[44] and gratings[45], photodetectors[46] including the ultrafast detection[47], as well as neuromorphic synapses[48] including the polarization sensitive ones[49].

Further development of the latter nanoscale devices allowing an operation at practical conditions calls for an establishment of a framework providing appropriate functionality at cryogenic and elevated temperatures, especially for, but not limited to applications in integrated hybrid nanophotonics, where an explicit insight of a variety of temperature-dependent properties, such as for example, the dielectric permittivity function is essential. The temperature dependent optical dispersion investigations of traditional Si or Ge semiconductors [50–53] and few grown TMD monolayers[54–57] propose that a consideration of conventional thermo-optic corrections in Vis (up to ~1100 nm NIR) spectral region is insufficient as the latter may evolve nonlinearly owing to the presence of excitonic contributions, though the treatment of dielectric permittivity of ternary TMDs or other vdW crystals as explicit functions of temperature is still essential.

In this manuscript, we present the dielectric permittivity functions of uniaxial TMD ternary MoSSe and WSSe in Vis spectral region partially covering NIR in a wide temperature range, 80–670 K. Our studies reveal strong room temperature *A*–*D* excitonic signatures emerging in spectral windows provided by their parent binary TMDs ($MoS_2$, $MoSe_2$, $WS_2$ and $WSe_2$) with high-refractive index (> 4) extinctionless NIR spectral tails uttered within the latter windows. We further demonstrate that the temperature dependencies of their dielectric permittivity functions in Vis spectral region are dictated by Varshni's relations that among others instigate spectral blue (red) shifts of excitonic resonances with temperature decrement (increment) tuning their in-plane refractive indices at respective NIR spectral tails. Furthermore, to elucidate the impact of temperature induced optical dispersion fine variation, leveraging the obtained dielectric permittivity functions of ternary WSSe, we design ultrathin plano-convex NIR nanophotonic lenses displaying continuous temperature-driven performance characteristics alterations.

## Results

### Room temperature dielectric functions of ternary and parent binary TMDs

Fig.1 demonstrates the dielectric permittivity functions of uniaxial ternary MoSSe and WSSe in framework of their parent binary crystals obtained by spectroscopic micro-ellipsometry (see Methods) studies from micro-mechanically cleaved samples. Here, ellipsometric parameters $\Psi$ and $\Delta$ were acquired at different angles of incidences (AoI) in the spectral region of 360–1000 nm for each TMD sample (see Methods and Figs. S1-S6) to provide highly accurate room temperature reference data. Notably, the obtained results for the parent TMD $MoS_2$, $WS_2$, $MoSe_2$ and $WSe_2$ bulk micro-crystals are in good agreement with previously reported data[18] reproducing the characteristic *A*, *B*, *C* and *D* excitonic absorption features in the imaginary parts of dielectric permittivity functions with a high exactitude (see Fig. 1 (b) and (e)). These optical transitions are governed by strongly bound excitons of Wannier–Mott type, modified by reduced dielectric screening[58,59]. The latter were obtained by introduction of Tauc–Lorentz oscillator-based modelling for description of in-plane components of dielectric permittivity functions with their imaginary parts defined as

$$\mathrm{Im}[\varepsilon(E)] = \frac{AE_o\Gamma(E-E_g)^2}{(E^2-E_o^2)^2+\Gamma^2E^2}\frac{1}{E}, E > E_g \quad (1)$$

where $E_o$ is the oscillator energy, $E_g$ is the optical bandgap, $\Gamma$ is the broadening parameter associated with the finite lifetime of the excited state often limited by the scattering processes, and $A$ is the oscillator's strength. The real parts of dielectric permittivity functions were then obtained by Kramers–Kronig relations according to

$$\mathrm{Re}[\varepsilon(E)] = \frac{2}{\pi} P \int_{E_g}^{\infty} \frac{\xi\varepsilon_2(\xi)}{\xi^2-E^2}d\xi, \quad (2)$$

where $P$ denotes Cauchy principal value of the integral (further details are provided in Methods).

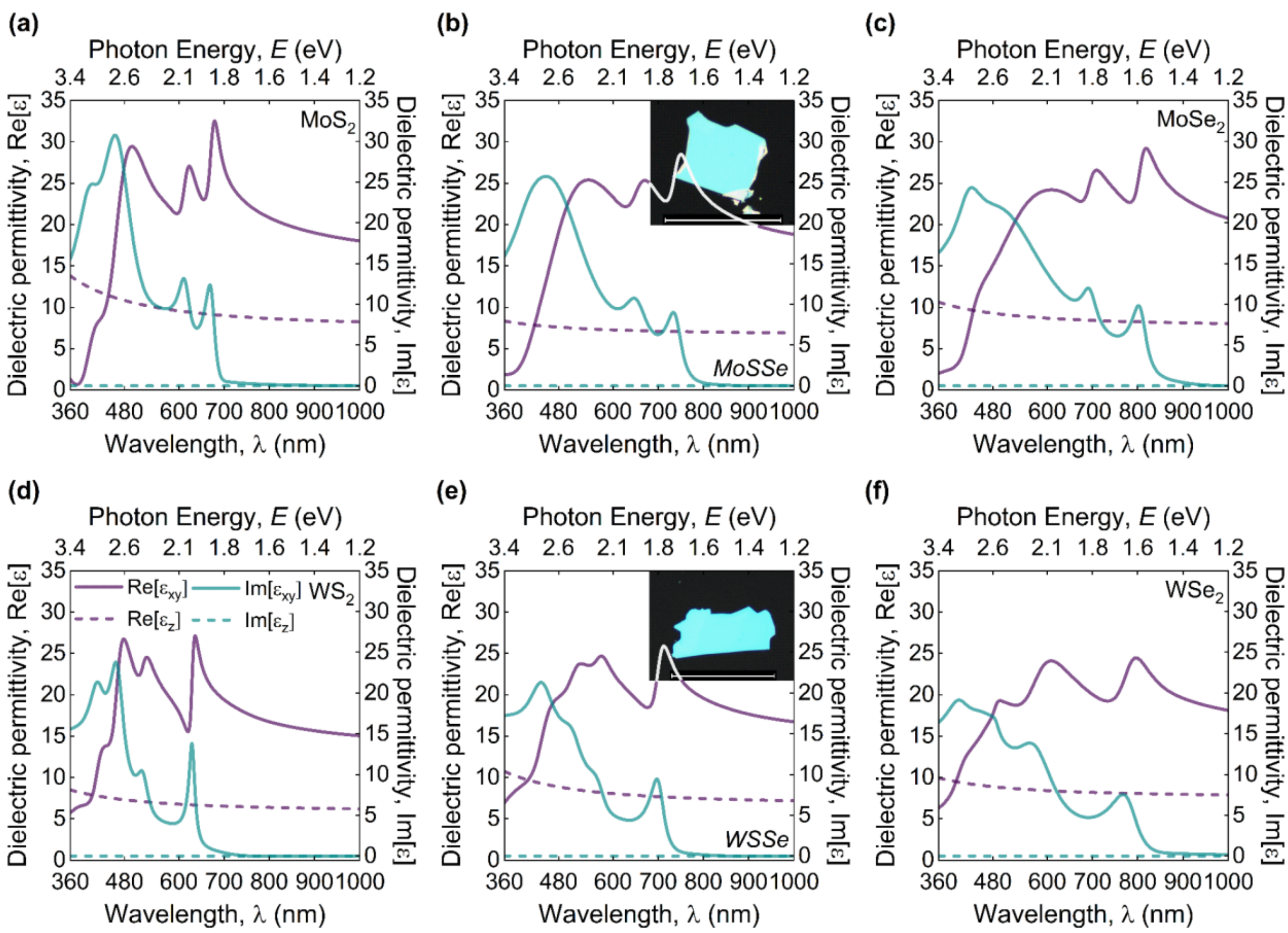

**Figure 1|Dielectric functions of uniaxial ternary MoSSe and WSSe in framework of their parent binary micro-crystals at room temperature. a-f,** Real (purple) and imaginary (cyan) parts of dielectric permittivity functions acquired by spectroscopic micro-ellipsometry (see Methods) from micro-mechanically cleaved (a) $MoS_2$, (b) MoSSe, (c) $MoSe_2$, (d) $WS_2$, (e) WSSe and (f) $WSe_2$ TMD samples in 360–1000 nm spectral region. The insets in (b) and (c) demonstrate 50 X optical micrographs of representative 62 nm and 68 nm thick MoSSe and WSSe samples utilized in ellipsometric studies. The scale bar corresponds to 75 μm. Solid curves in represent the in-plane dielectric permittivity components, Re[$\varepsilon_{xy}$] and Im[$\varepsilon_{xy}$], while dashed curves represent the out-of-plane components, Re[$\varepsilon_z$] and Im[$\varepsilon_{xy}$]. The extracted room temperature optical constants and models of ternary MoSSe and WSSe are presented in Supplementary Note 1.

The resulting real and imaginary parts of dielectric permittivity functions display trends resembling those of their parent TMDs, including the super-mossian characteristics[60]. An excitonic footprint of the same nature is as well observed in the ternary MoSSe (and WSSe) with distinct absorption peaks emerging in imaginary parts of the dielectric permittivity functions at 1.67 (and 1.76) eV, 1.90 (and 2.17) eV, 2.58 (and 2.41) eV, and 2.97 (and 2.79) eV. Notably, the two high-energy peaks of the former ternary samples appear merged into a singular feature (see Fig. S7). A comparison of the robust *A*-excitonic peaks of $MoS_2$ (1.84 eV) and $MoSe_2$ (1.53 eV), as well as $WS_2$ (1.97 eV) and $WSe_2$ (1.58 eV), leads to an evident conclusion indicating that the excitonic energies are governed by chalcogen elements, with sulphur-based TMDs exhibiting high energy sharper peaks of reduced broadening. In the case of our ternary samples, the latter peaks emerge at 1.76 eV (WSSe) and at 1.67 eV (MoSSe) representing intermediate energies between their parent binary TMDs, and hence, indirectly inferring a stoichiometric composition of $S_{0.5}Se_{0.5}$.

Furthermore, the absorption onsets, *i. e.*, the inceptions of optically transparent NIR spectral regions, in our ternary samples also emerge within the windows provided by their parent TMDs. Notably, sulphur-based TMDs with absorption edges occurring at shorter wavelengths (see Fig. 1 (a) and (d)) display comparatively lower, but still high in-plane refractive indices at NIR spectral tail (see Fig. S8). In contrast, selenium-based ones, whose absorption edges incept at longer wavelengths (see Fig. 1 (c) and (f)), exhibit higher in-plane refractive indices at same NIR spectral tail (see Fig. S8). This behaviour is associated with broader excitonic transitions, which extend their influence further into the transparent spectral region. Our ternary samples follow this trend granting simultaneous access to the negligible losses ($k_{xy} \leq 0.01$) at NIR spectral region above 850 nm (and 822 nm) wavelengths and high in-plane refractive indices of $n_{xy} = 4.63$ (and $n_{xy} = 4.36$) in the case of MoSSe (and WSSe) compared to lower refractive index sulphur based or lossy, and, therefore, inaccessible, selenium-based parent TMDs (see Fig. S8 for more details). In addition, to validate a high accuracy of obtained room temperature reference data, we further employed Vis micro-reflectance spectroscopy of our ternary MoSSe and WSSe (and their parent binary micro-crystals) placed on a variety of substrates. The measured and modelled spectra shown in Figs. S9-S14 display quite good agreement with a small overall fitting error of ≤ 3 %.

### The temperature dependence of dielectric functions of ternary TMDs

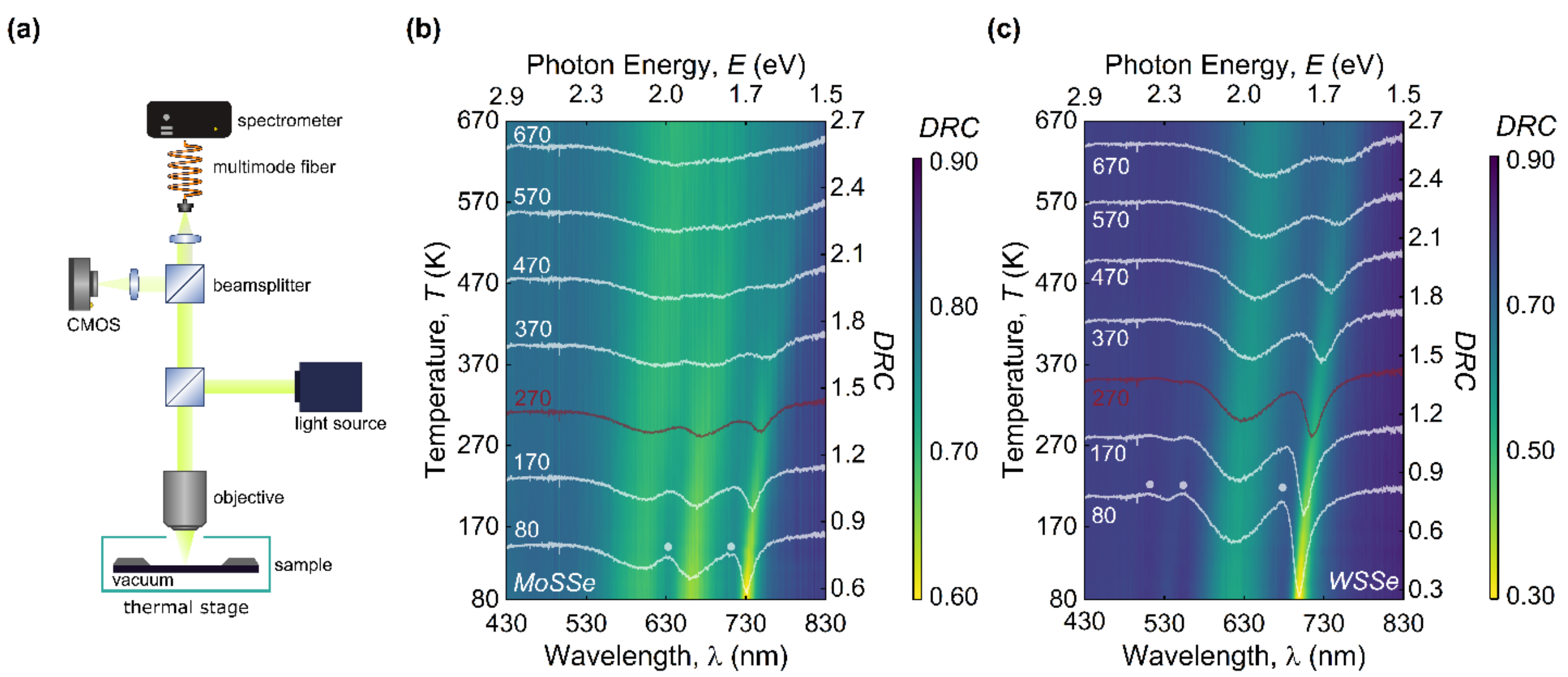

**Figure 2 | Differential reflectance contrast (*DRC*) of uniaxial MoSSe and WSSe ternary TMDs. a,** Schematic illustration of experimental setup employed for temperature-dependent studies of the optical response. **b,** and **c,** *DRC* contour plots as function of wavelength (energy) and temperature for micro-mechanically cleaved representative MoSSe (b) and WSSe (c) samples placed onto fused silica substrates and measured in 430–830 nm spectral region at 80–670 K temperature range. White curves indicate selected *DRC* spectra at several temperatures to visualize the evolution of the critical points with temperature. An offset of 0.3 is introduced for clarity. The red curves highlight *DRC* spectra measured at 270 K (~0 °C). White dots mark the critical points at 80 K corresponding to excitonic transitions. The spectral maps were acquired through the merging of the spectra in two spectral regions (see Methods and Supplementary Note 2).

The temperature-dependence of in-plane components of dielectric permittivity functions of ternary MoSSe and WSSe were extracted from micro-reflectance spectroscopy acquired at normal incidence using a custom designed setup schematically illustrated in Fig. 2(a). The latter comprises an upright optical microscope with a focusing system specifically devised to collect reflected spectra from micro-mechanically cleaved samples integrated into an optical thermal stage. Here, MoSSe and WSSe samples were placed onto fused silica substrates and mounted into an optical thermal stage providing stable isolation from ambient environment, preventing condensation at low temperatures (see Methods for further details). Fig. 2(b) and 2(c) show *DRC* spectral maps acquired over temperature range of 80–670 K with the highlighted curves indicating selected temperature responses. To ensure high accuracy of the obtained spectra, the maps comprise measurements merged from two spectral regions (see Methods and Figs. S15-S16 for more details).

The excitonic absorption features of our ternary MoSSe and WSSe were directly observed in *DRC* spectra and identified as critical points. In ternary MoSSe, *A* and *B* excitonic states are resolved by a naked eye, while in ternary WSSe, one may also clearly observe robust *C* state boosted critical points (see white dots in Fig.2 (b) and (c)). At shorter wavelengths (higher energies), the absorption and reflectance exhibit high values reducing the contrast of high-energy transitions and making them less distinguishable within studied spectral region. Thus, the remaining excitonic state parameters (invisible to a naked eye but still playing an important role in the extraction of imaginary part of dielectric permittivity function) were implemented from the spectroscopic micro-ellipsometry data into a room temperature reflectance spectra framework (see Fig. S17). Fig. 2 (b) and (c) demonstrate the systematic temperature dependence, where the corresponding critical points blue-shift (red-shift) toward shorter wavelengths with temperature decrement (increment). Importantly, this behavior is characterized solely by spectral shifts of existing critical points suggesting main effect to be governed by the modification of Tauc–Lorentz oscillator parameters rather than the emergence of new optical transitions within studied spectral region (down to 80 K). Therefore, we used the same parametrized model to describe the excitonic absorption features by tracking an evolution of oscillator parameters as function of temperature in a simultaneous fit of temperature dependent *DRC* spectra (see Methods and Figs. S18-S19). Here, the oscillator energies were described by Varshni's relation $E_o(T) = E_o(0) - \alpha T^2/(\beta+T)$, where $E_o(0)$ is the transition energy at 0 K, and $\alpha$ and $\beta$ are constants describing the temperature-induced shifts of optical bandgaps[61] and $\Gamma(T) = \Gamma(0) + \gamma_{LO}/(exp(\Theta/T)-1)$ stands for broadening parameter, which arises from Bose–Einstein distribution of phonons, $\Gamma(0)$ represents the temperature-independent broadening, $\gamma_{LO}$ describes the coupling strength of electronic-longitudinal optical phonon interactions and $\Theta$ is the characteristic phonon temperature.

The extracted in-plane components of dielectric permittivity functions of the ternary TMD WSSe and MoSSe exhibit (see Fig. 3(a, b) and Fig. 3(e, f)) temperature evolution consistent with the imposed parametrization, following dependencies prescribed by Varshni's relations. The maps resolve multiple excitonic contributions labelled as *A*, *B*, *C* + *D* for ternary MoSSe and *A*-*D* for ternary WSSe, whose evolution with temperature is reflected in gradual spectral shifts and broadening diminishment (see Fig. 3(c, d) and Fig. 3(g, h)) with temperature decrement. The fit-constrained parameters obtained from spectroscopic micro-ellipsometry at room temperature demonstrate good agreement with data extracted from micro-reflectance spectroscopy (see the red markers).

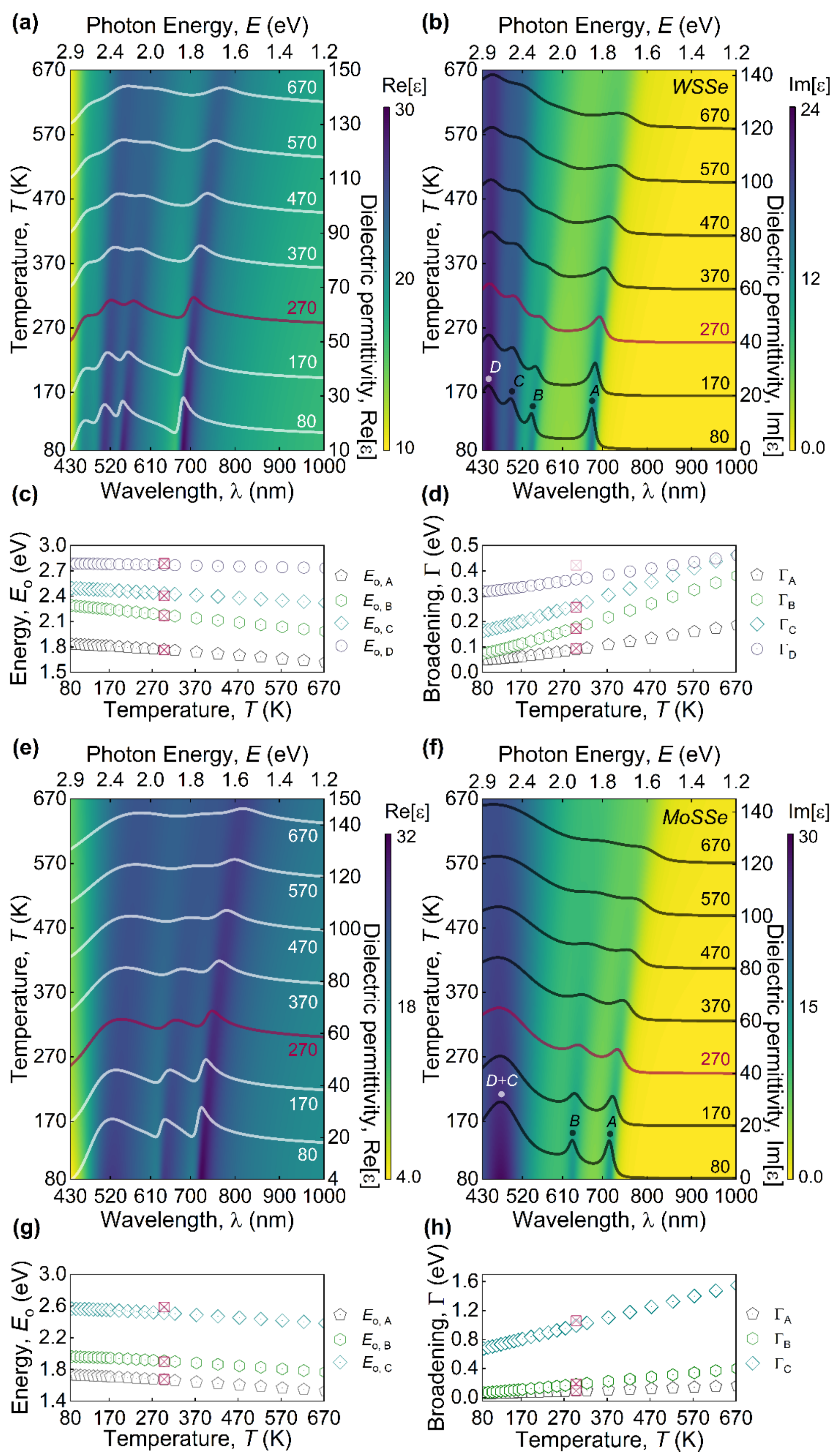

**Figure 3| Temperature dependence of in-plane dielectric permittivity functions of uniaxial MoSSe and WSSe.** Real **a,** and **e,** and imaginary **b,** and **f,** parts of dielectric permittivity components, Re[$\varepsilon_{xy}$] and Im[$\varepsilon_{xy}$], presented by contour plots as function of wavelength (energy) and temperature for ternary WSSe (a, b) and MoSSe (e, f). White curves indicate the selected dielectric functions

at several temperatures to visualize the evolution of excitonic states with temperature variation. An offset of 0.2 is introduced for clarity. The red curves highlight the dielectric functions at 270 K (~0 °C). Coloured dots mark the excitonic transition peaks in imaginary part of dielectric functions at 80 K. The evaluation of **c,** and **g,** Tauc–Lorentz oscillator peaks, $E_o$, and **d,** and **h,** broadening factor, $\Gamma$, with the variation of temperature for ternary (c, d) WSSe and (g, h) MoSSe. Red markers indicate room temperature values obtained from spectroscopic micro-ellipsometry studies used as initial parameters during the fitting procedure. The extracted low temperature optical constants and models of ternary MoSSe and WSSe are presented in Supplementary Note 3.

While all excitonic states follow the same functional dependence, their parameters reveal a presence of non -uniformity suggesting various extents of electron–phonon interactions. The latter is evidenced by Varshni's coefficients (see Tables S5-S6) governing the energy shifts and phonon coupling parameters controlling the broadening, which may further provide insights into the nature of excitonic transitions. For example, *A*-excitons in ternary MoSSe and WSSe exhibit minimal broadening indicating that they remain the most robust optical transition across entire studied temperature range. Furthermore, a phonon-induced broadening is noticeably stronger in ternary WSSe suggesting enhanced electron–phonon coupling in comparison to the former. The high-energy excitonic states in our ternary compounds display significantly larger temperature-dependent broadening, indicating increased scattering and reduced coherence. The optical bandgaps reveal pronounced temperature-induced red-shift (blue-shift) of transparency edge with an increment (decrement) of temperature. Specifically, for ternary WSSe, the bandgap shifts from 1.72 eV (721 nm) to 1.50 eV (827 nm), while for ternary MoSSe, it shifts from 1.61 eV (770 nm) to 1.41 eV (879 nm) over the temperature range of 80–670 K. Notably, as the transparency edge shifts to longer wavelengths with an increment of temperature, the real part of dielectric permittivity function increases in accordance with Kramers–Kronig relations. This results in an increase of the in-plane refractive index in transparent NIR spectral region. For example, at 1000 nm wavelength, the real part of dielectric permittivity increases from 18.14 (and 16.55) to 20.8 (and 18.34) for ternary MoSSe (and WSSe) allowing a lossless window of the in-plane refractive index tunability (see Fig. S20-S21) from 4.259 (and 4.069) to 4.561 (and 4.282).

**The effect of temperature in ultrathin NIR photonic lenses based on ternary WSSe**

TMDs exhibit particularly large refractive indices across Vis and NIR spectral regions allowing their few layers to generate substantial optical path lengths (*OPL*) through strong interfacial reflections, enabling pronounced phase modulation in ultrathin geometries in comparison with graphene[62]. This enables wavefront shaping in flat nanophotonic components paving the way to ultrathin lenses[63] with significantly improved focusing efficiencies when structured into Fresnel's zone plates[64]. Following these advances, we propose a design of plano–convex ultrathin NIR lenses based on a suspended ternary WSSe operating in the transmission regime (see Fig. 4(a)) and examine the effect of temperature on its optical performance. Our lens relies on a phase accumulation induced by a spatially varying thickness profile and is designed for operation at a wavelength of 925 nm, where ternary WSSe remains infirmly absorbing over the considered temperature range. Here, we first compute OPL using a transfer-matrix approach (see Methods) for a selected optimized geometry with a central thickness of 16 layers and an edge thickness of 3 layers (total height of 13 layers). The corresponding thickness and *OPL* profiles are shown in Fig. 4(b). Next, to assess the temperature induced effects, we utilize the obtained dielectric permittivity function of ternary WSSe recalculating *OPL* profiles, which are shown in Fig. 4(c). Our studies display a continuous shift of the focal length from 835.4 μm at 80 K to 773 μm at 670 K emphasizing a constant temperature-driven modification of lens's performance. This shows that an ultrathin suspended ternary WSSe lenses not only enable robust wavefront engineering but also exhibit continuous temperature-dependent performance characteristics tuning, which should be accounted in designing of next-generation thermo-driven nanophotonic components.

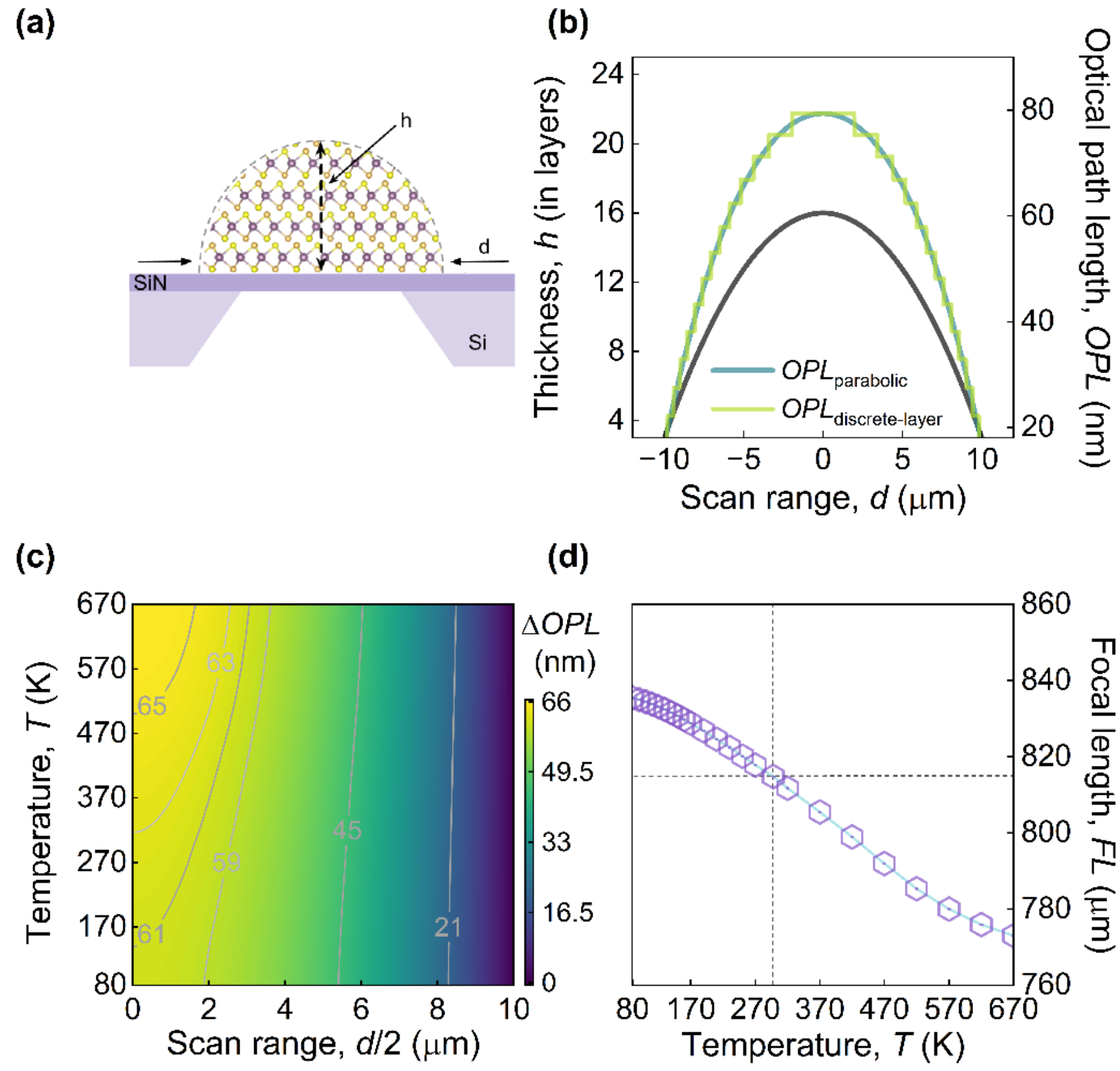

**Figure 4|Focal length temperature dependence in ultrathin NIR photonic lenses based on suspended ternary WSSe. a,** Schematic illustration of a plano–convex transmission lens based on WSSe optical layer suspended over SiN membrane. **b,** Thickness profile of the lens geometry and its *OPL* distribution at 925 nm wavelength. Light green curve shows discrete layer *OPL*s. The central thickness of the lens corresponds to 10.72 nm, while the edge thickness to 2.01 nm. **c,** The temperature-dependent contour plot of edge-relative *OPL* distribution across the scan range (see Methods). **d,** Calculated focal length, *FL*, as a function of temperature in 80–670 K range. The dashed lines highlight the focal length at room temperature.

## Discussion and Conclusion

The extraction of dielectric permittivity function temperature evolution of uniaxial ternary MoSSe and WSSe was enabled by a hybrid approach combining spectroscopic micro-ellipsometry and temperature-dependent micro-reflectance spectroscopy with modeling, reconstructing their in-plane components across a wide range of temperatures, 80–670 K. The spectroscopic micro-ellipsometry provides reliable determination of room temperature dielectric permittivity functions, but exhibits significant challenges when extended to cryogenic temperatures for yet supreme quality micro-mechanically cleaved samples possessing limited lateral sizes.[65] In contrast, micro-reflectance spectroscopy alone, though allowing much swifter capabilities, is ineffective in providing accurate solutions for the dielectric permittivity function determination mostly due to the presence of multiple overlapping excitonic or other interband transitions, which often introduce strong correlations between oscillator parameters and nanoscale sample thicknesses. Here, by using ellipsometric parameters as a physically constrained starting point and simultaneously fitting the temperature-dependent reflectance spectra, we establish a unique approach that enables consistent and reliable determination of dielectric permittivity as a function of temperature in a wide spectral region. Our technique can be potentially extended to a broader class of TMD (or other vdW) crystals including biaxial ones paving the way to creation of next-generation anisotropic photonic cryo or elevated temperature components introducing limitations only for those with high optical absorptions that are at the same time accompanied with high refractive indices at the certain wavelengths. Our findings present the temperature evolution of *A*–*D* excitonic features in uniaxial ternary MoSSe and WSSe at Vis spectral region that further dictate the temperature dependent properties of their high-refractive index (> 4) lossless NIR spectral tails. Furthermore, using obtained dielectric permittivity functions of ternary WSSe, we design ultrathin plano-convex NIR photonic lenses, which display continuous temperature-driven performance alterations based on temperature induced optical dispersion variation at the respective NIR spectral tail.

## Methods

**Sample preparation.** Hexagonal $MoS_2$, $MoSe_2$, $WS_2$, $WSe_2$, MoSSe and WSSe crystals were purchased from 2D Semiconductors and HQ graphene and micro-mechanically cleaved onto standard Si, $Si/SiO_2$ and fused silica substrates at room temperature using commercial scotch tapes from Nitto Denko Corporation. Prior to micro-mechanical cleavage, the substrates were subsequently cleaned in acetone, isopropanol and deionized water followed by air plasma treatment to remove residual surface contaminants and improve the adhesion.

**Room temperature spectroscopic micro-ellipsometry.** Spectroscopic micro-ellipsometry was performed at room temperature over 360-1000 nm spectral region using Park Systems Accurion EP4 rotating-compensator imaging ellipsometer. All modelling was carried out using EP4Model software. Multiple $MoS_2$, $MoSe_2$, $WS_2$, $WSe_2$ and ternary MoSSe and WSSe samples of varying thicknesses and across various substrates (Si, $Si/SiO_2$ and fused silica) were measured and analysed to ensure the reliability and improve the accuracy of obtained data. Spectra of the ellipsometric parameters $\Psi$ and $\Delta$ were acquired at AoIs of 45°, 55°, 60°, 65° and 70°. Variable-angle spectroscopic ellipsometry (VASE) measurements were also performed to minimize correlations between fitting parameters in the anisotropic model. An anisotropic optical model with distinct in-plane and out-of-plane components was constructed and fitted. The in-plane component of each TMD and their ternary sample was parameterized using Tauc-Lorentz oscillators. For the description of out-of-plane component, Cauchy model was employed in all the cases. The total in-plane dielectric permittivity was represented as the sum of all oscillators by

$$\varepsilon(E) = \varepsilon_\infty + \sum_{i=1}^{N} TL_i(E) \quad (3)$$

The dielectric permittivity functions and corresponding optical constants of $MoS_2$, $MoSe_2$, $WS_2$ and $WSe_2$ and ternary MoSSe and WSSe were obtained by minimizing the mean squared error of the model, examining parameter correlation matrix and accounting for depolarization effects arising from finite source bandwidth (see Supplementary Note 1 for details).

**Temperature dependent micro-reflectance spectroscopy.** Micro-reflectance spectra of ternary MoSSe and WSSe were measured under a normal incidence using Leica DM6M upright optical microscope equipped with LED and halogen light sources covering 430–830 nm spectral region. A CRYO600-190 optical cryo-stage, equipped with a $LN_2$ cooling system and a vacuum pump, was used to perform measurements in 80–670 K temperature range. The spectra presented in Fig. 2 were obtained by combining measurements from both light sources to minimize the noise across the spectral region. The LED source was used at photon energies above 2.25 eV for ternary WSSe (and 2.15 eV for ternary MoSSe), while halogen source was used below this energy providing improved signal-*to*-noise performance in the respective spectral regions (see Figs. S15-S16). Spectral acquisition was performed using Thorlabs CCS100/M (350–700 nm, for LED illumination) and CCS200/M (200–1000 nm, for halogen illumination) spectrometers, coupled *via* 105 µm core-diameter optical fibre. In combination with a 20 X objective[66], the latter configuration enabled spectral measurements from regions with lateral dimensions of approximately 10 µm. Due to the infinitesimal temperature coefficient of the refractive index of fused silica[67] compared to ternary MoSSe and WSSe, the substrate reflectance was measured at 270 K (0 °C) and used as a reference for all the temperatures. The differential reflectance contrast was calculated as DRC = $(J_s - J_{sub})/(J_s + J_{sub})$, where $J_s$ and $J_{sub}$ are the reflected intensities from the sample and the substrate. Measurements were performed at 24 different temperatures (see Supplementary Note 2). At the low-temperatures (80–170 K), a 10 K step was used, intermediate 190–270 K temperatures were measured with 20 K step, followed by measurements at 298 K (room temperature). At higher temperatures (320–670 K), a 50 K step was applied. The dielectric permittivity functions were obtained from *DRC* spectra by modelling the imaginary parts using Tauc–Lorentz oscillator formalism with initial parameters derived from spectroscopic micro-ellipsometry at room temperature. The real part of dielectric permittivity function was calculated from the imaginary part using fast Kramers–Kronig transformation, including a high-energy tail correction. The optical bandgap, $E_g$, was defined relative to oscillator's energy, while the oscillator amplitudes were treated as free fitting parameters. The model parameters were determined by minimizing the residual among calculated and measured spectra using a trust-region reflective nonlinear least-squares algorithm. The presented dielectric permittivity functions of ternary MoSSe and WSSe were extrapolated up to 1000 nm from 830 nm using the fitted model parameters presented in Supplementary Note 3.

**Calculations of ultrathin plano-convex lenses.** The lensing of the suspended WSSe was modelled by utilizing a transmission-based formalism for an *air*-TMD-*air* structure combined with a phase-accumulation approach, consistent with methods earlier used for the designing of atomically thin optical elements[63]. The optical path length, *OPL*, was calculated as function of thickness using transfer-matrix formalism for *air*-TMD-*air* structure. The complex transmission coefficient $t(d, \lambda)$ was referenced to an equal-thickness air path with a phase difference defined as $\Delta\varphi(d, \lambda) = arg(t_{sample}(d, \lambda)/t_{air}(d, \lambda))$ yielding $OPL(d, \lambda) = -\lambda\Delta\varphi(d, \lambda)/2\pi$. A continuous OPL–thickness relation was constructed and used to map spatial thickness profiles into phase profiles. The lens geometry was described by a rotationally symmetric convex profile. For a lens of radius $r$, edge thickness $t_{edge}$ and central thickness $t_c$, the thickness distribution was defined as $t(x) = \{t_{edge} + (t_c - t_{edge})(1 - x^2/r^2)$, $|x| \leq r$ and $t_e$, $|x| > r$, where $x$ is the lateral position across the lens. The spatial *OPL* profile ($OPL(x)$) was obtained by mapping the thickness profile $t(x)$ onto the precomputed $OPL(d)$ relation. The focusing was evaluated based on phase accumulation across the aperture, following the phase-front engineering approach. The transmitted phase profile was then defined as $\varphi(x) = -\lambda/2\pi[OPL(x) - OPL_{edge}]$, where the subtraction of $OPL_{edge}$ removed the global phase offset. The on-axis field at axial distance $h$ from the lens was computed as $E(h) \propto \int_{-r}^{r} exp\left(i\varphi(x)\right) exp(-\mathrm{i}\lambda\sqrt{h^2 + x^2}/2\pi) dx$. The axial intensity distribution was obtained as $I(h) = |E(h)|^2$ and the focal length, *FL*, was defined as the position of its maximum. In all calculations, the monolayer thickness of WSSe was taken as 0.67 nm. The lens geometries were explored by scanning the parameter space defined by edge thickness, central thickness and lens radius over predefined ranges. For each candidate geometry, the corresponding *OPL* profile and focal length were calculated using the same propagation model. In the focal-length-targeted optimization, candidates were selected by minimizing the relative deviation from a target while enforcing a minimum centre-*to*-edge *OPL* contrast. In a second approach, the computed *OPL* profile was compared with an ideal quadratic dependence corresponding to an ideal aberration-free lens phase profile. The deviations from this quadratic behaviour were quantified using a normalized root-mean-square error together with transmission-based efficiency metric. The temperature dependence of *OPL* and focal length for the fixed lens geometry were evaluated by using obtained temperature-dependent dielectric constant dataset of ternary WSSe.